\documentclass[reprint,nofootinbib,superscriptaddress,amsmath,amssymb,aps]{revtex4-1}
\usepackage{graphicx}
\usepackage{rotating}
\usepackage{dcolumn}
\usepackage{bm}
\usepackage{color}
\usepackage{mathptmx, textcomp}
\usepackage[latin1]{inputenc}
\usepackage{braket}
\usepackage[colorlinks,linkcolor=magenta,anchorcolor=cyan,citecolor=blue]{hyperref}
\usepackage[multiple]{footmisc}

\def\be{\begin{equation}}
\def\ee{\end{equation}}
\def\ba{\begin{eqnarray}}
\def\ea{\end{eqnarray}}

\newcommand{\eq}[1]{(\ref{#1})}

      \def\p {\pi}     \def\g {\gamma}    \def\l {\lambda}   \def\c {\chi} \def\b {\beta}   \def\pd {\partial}\def\p {\pi}   \def \e { \varepsilon}
              \def\grad{\nabla}\def\.{\cdot}
\def\math {\mathcal}
\begin{document}

\title{Investigating the Linearized Second Law in Horndeski Gravity}
\author{Xin-Yang Wang}
\email{xinyang\_wang@foxmail.com}
\affiliation{College of Education for the Future, Beijing Normal University, Zhuhai 519087, China}
\author{Jie Jiang}
\email{Corresponding author. jiejiang@mail.bnu.edu.cn}
\affiliation{Department of Physics, Beijing Normal University, Beijing 100875, China\label{addr2}}
\date{\today}

\begin{abstract}
Since the entropy of stationary black holes in Horndeski gravity will be modified by the non-minimally coupling scalar field, a significant issue of whether the Wald entropy still obeys the linearized second law of black hole thermodynamics can be proposed. To investigate this issue, a physical process that the black hole in Horndeski gravity is perturbed by the accreting matter fields and finally settles down to a stationary state is considered. According to the two assumptions that there is a regular bifurcation surface in the background spacetime and that the matter fields always satisfy the null energy condition, one can show that the Wald entropy monotonically increases along the future event horizon under the linear order approximation without any specific expression of the metric. It illustrates that the Wald entropy of black holes in Horndeski gravitational theory still obeys the requirement of the linearized second law. Our work strengthens the physical explanation of Wald entropy in Horndeski gravity and takes a step towards studying the area increase theorem in the gravitational theories with non-minimal coupled matter fields.
\end{abstract}
\maketitle

\section{Introduction}
In general relativity, the area theorem of black holes which states that the area of the event horizon will never decrease during physical processes has been demonstrated by Hawking \cite{Hawking:1971tu}. From the theorem, Bekenstien \cite{Bekenstein:1973ur} proposed a conjecture firstly that the area of the event horizon may be identical to the entropy of black holes. Subsequently, Hawking \cite{Hawking:1974sw} first proved the temperature of black holes is proportional to the surface gravity, while the entropy can be written as $S_\text{BH} = A / 4$, which is called Bekenstein-Hawking entropy. It implies that black holes can be regarded as an adiabatic system in the thermodynamics. Furthermore, the four laws of mechanics for black holes have been constructed \cite{Bekenstein:1972tm, Bardeen:1973gs}. The two profound laws in the four laws of mechanics are the first and the second law. The first law can be expressed as $dE = T dS$ for stationary black holes, where $E$ is the Killing energy which can be measured in corotating coordinates, $T$ is Hawking temperature, and $S$ is the entropy of black holes. The second law is stated that the entropy of the black hole will increase irreversibly. For the generalized second law of black holes, the entropy does not only contain Bekenstein-Hawking entropy, it demands that the sum of the entropies of the horizon and the matter outside black holes, $S = S_\text{BH} + S_\text{out}$, will always increase with physical processes \cite{Bekenstein:1974ax}.

However, for any generally covariant theory of gravity, a question of whether black holes in the corresponding gravitational theory can be considered as a thermodynamic system as well will be naturally raised. From this question, the modified first law for the equilibrium state of black holes in any diffeomorphism invariant gravitational theory has been established based on the Iyer-Wald formalism \cite{Wald:1993nt, Iyer:1994ys}. In the modified first law, the entropy is called the Wald entropy, and it is no longer proportional to the area of the event horizon. If we recognize that the gravitational dynamics of black holes strongly connects to the thermodynamics for any diffeomorphism invariant gravitational theory, the validity of the four laws of mechanics should be examined again, especially the second law. Following this perspective, the generalized second law for other kinds of black holes has been investigated. Using the method of the field redefinition, it is shown that the Wald entropy for black holes in the $f(R)$ gravity obeys the requirement of the second law \cite{Jacobson:1993vj, Jacobson:1995uq}. However, there is still existing the situation that the entropy of black hole does not obey the second law of black hole thermodynamics. For the Lagrangian which contains higher-order curvature terms, the second law can be violated in the case of the two black holes merging \cite{Sarkar:2010xp}. However, Ref. \cite{Bhattacharjee:2015yaa} has been argued that if the quantum corrections of the black hole is involved, it is sufficient to examine the second law under the first order approximation in an adiabatic process. When only concerning the first-order approximation, the second law of black holes in Gauss-Bonnet and Lovelock gravitational theories have been investigated  \cite{Chatterjee:2011wj, Kolekar:2012tq}. Subsequently, a general proof of the linearized second law in higher curvature gravity has been proposed by Wall \cite{Wall:2015raa}, and the expression of the entropy which satisfies the linearized second of black holes in $F($Riemann$)$ gravity has been obtained.

Although the general relativity is the most successful theory to describe the interaction of the gravity, it cannot provide a satisfactory interpretation of some cosmological phenomena, such as the origin of the early Universe, the accelerated expansion, and the present of the dark matter and the dark energy. However, the scalar field is considered as a suitable candidate to solve these phenomena \cite{Brans:1961sx}. From this perspective, several cosmological models which contain the correction of scalar fields have been proposed \cite{Bertacca:2007ux, Chung:2007vz, delaMacorra:2007beq, Saridakis:2010mf}. Recently, the Horndeski gravitational theory, which contains a non-minimally coupled axionic scalar field, has received much attention through their application to cosmology in Galileon theories \cite{Nicolis:2008in}. Although the Lagrangian involves more than two derivatives, the field equations and the stress-energy tensor involve no higher than second derivatives of the fields \cite{Cisterna:2014nua, Jiang:2017imk}. On the other hand, since the scalar field couples to the Riemann curvature in Horndeski gravity, the formalism of the Wald entropy will be influenced by the scalar field. Therefore, it is natural to ask whether the Wald entropy of black holes in Horndeski gravity still obeys the second law of black hole thermodynamics. Following the line of through in Ref. \cite{Kolekar:2012tq}, we would like to examine the linearized second law of the Wald entropy in Horndeski gravity under the linear-order approximation.

The organization of the paper is as follows. In Sec. \ref{sec2}, we introduce the Horndeski gravitational theory and the definition of the Wald entropy. In Sec. \ref{sec3}, considering the matter fields perturbation, while according to the two assumptions that a regular bifurcation surface exists in the background spacetime and that the matter fields satisfy the null energy condition, we investigate whether the Wald entropy for black holes in Horndeski gravitational theory obeys the second law under the linear order approximation. The paper ends with discussions and conclusions in Sec. \ref{sec4}.

\section{Horndeski gravitational theory and Wald entropy}\label{sec2}

We consider the $(n+2)$-dimensional Horndeski gravitational theory minimally coupling to some additional matter fields which satisfy the null energy condition. The action is given by
\ba\begin{aligned}
I=I_\text{Horn}+I_\text{mt}
\end{aligned}\ea
with
\begin{equation}\label{action}
	I_\text{Horn}= \frac{1}{16 \pi} \int d^{n+2} x \sqrt{-g} \left[R- \frac{1}{2} \left(\beta g^{ab} - \alpha G^{ab} \right)  \nabla_a \chi \nabla_b \chi \right]\,,
\end{equation}
where $I_{\text{mt}}$ is the action of the additional matter fields, $G_{ab}=R_{ab} -(1/2)Rg_{ab}$ is the Einstein tensor, $\chi$ represents the scalar field, while $\alpha$ and $\beta$ are the coupling constants between the gravity and the scalar field. The equation of motion of the gravitational part is given by
\ba\begin{aligned}
H_{ab}=8\p \left(T_{ab}^{(\b)}+T_{ab}^\text{mt}\right)\,,
\end{aligned}\ea
in which $T_{ab}^\text{mt}$ is the stress-energy tensor of the additional matter fields, and we have denoted
\ba
\begin{aligned}
T_{ab}^{(\b)}=\frac{\b}{16\p}\left(\grad_a\c\grad_b\c-\frac{1}{2}g_{ab}\grad_c\c\grad^c\c\right)\\
\end{aligned}\ea
and
\begin{equation}
	\begin{aligned}
		&H_{ab} = G_{ab} - \frac{\alpha}{2} \left[\frac{R}{2} \nabla_a \chi \nabla_b \chi -2 \nabla_c \chi \nabla_{(a} \chi R_{b)}^{\ c} + \nabla_a \nabla_b \chi \nabla^2 \chi \right. \\
		& \left. -R_{acbd} \nabla^c \chi \nabla^d \chi - \nabla _a \nabla^c \chi \nabla_b \nabla_c \chi +\frac{1}{2}G_{ab} \left(\nabla \chi\right)^2 \right] \\
		& - \frac{\alpha}{4}g_{ab} \left[\left(\nabla^c \nabla^d \chi\right) (\nabla_c \nabla_d \chi) - \left(\nabla^2 \chi \right)^2 + 2 R^{cd} \nabla_c\chi \nabla_d \chi \right]\,.
	\end{aligned}
\end{equation}
From the above expressions, we can see that $T_{ab}^{(\b)}$ is exactly the stress-energy tensor of the minimally coupled scalar field, while it is required that the scalar field should satisfy the null energy condition, i.e., $T_{ab}^{(\b)}k^ak^b\geq 0$ for any null vector along the future direction $k^a$. Therefore, the term which contains the minimally coupled scalar field can be collected into the additional matter fields, and the equation of motion can be rewritten as
\ba\begin{aligned}
H_{ab}=8\p T_{ab}\,,
\end{aligned}\ea
where
\ba\begin{aligned}
T_{ab}=T_{ab}^{(\b)}+T_{ab}^\text{mt}
\end{aligned}\ea
represents the total stress-energy tensor, and it is further required that the total stress-energy tensor should satisfy the null energy condition as well.

For stationary black holes, the entropy is given by the Wald entropy
\begin{equation}
	S = -2 \pi \int d^n y \sqrt{\gamma} \frac{\partial \mathcal{L}}{\partial R_{abcd}} \bm{\epsilon}_{ab} \bm{\epsilon}_{cd}\,,
\end{equation}
where $\mathcal{L}$ is the Lagrangian of the Horndeski gravitational theory, $\bm{\epsilon}_{ab}$ is the binormal on a specific slice of the event horizon, $\sqrt{\gamma}$ is the volume element of any slice of the horizon, and $y$ labels the transverse coordinate of the cross section of the horizon. According to the definition of the Wald entropy and Eq. (\ref{action}), the Wald entropy of black holes in Horndeski gravitational theory is obtained as
\begin{equation}
	S = \frac{1}{4} \int_s d^\text{n} y \sqrt{\gamma} \left(1- \frac{\alpha}{4} D_a\chi D^a \chi \right)\,,
\end{equation}
where the symbol $s$ represents the cross section of the event horizon, and $D_a X_{a_1a_2\cdots}= \gamma_{a}^{\ b}\g_{a_1}^{\ b_1}\g_{a_2 \cdots}^{\ b_2 \cdots} \nabla_b X_{b_1b_2\cdots}$ for any spatial tenor $X_{a_1a_2\cdots}$ is denoted as the spatial operator of the covariant derivative which is compatible with the induced metric $\gamma_{ab}$, i.e., $D_c \gamma_{ab} = 0$. According to the expression of the Wald entropy, we can see that the entropy of the stationary black hole is modified by the scalar field and is no longer proportional to the area of the event horizon. Furthermore, a meaningful question of whether the Wald entropy in Horndeski gravity still obeys the second law of the thermodynamics can be naturally proposed.

\section{Examining the linearized second law for black holes in Horndeski gravity}\label{sec3}
In the following, a slow accreting process that the matter fields pass through the event horizon to perturb the black hole is considered. For the process, it should be required that the spacetime geometry of the black hole finally settles down to a stationary state after the perturbation, and the matter fields satisfy the null energy condition as well. The event horizon is denoted as $\mathcal{H}$, which is a $n$-dimensional null hypersurface and can be generated by the null vector field $k^a = (\pd/\pd\l)^a$. If $\l$ can be chosen as an affine parameter, the null vector field $k^a$ obeys the geodesic equation $k^b \nabla_b k^a = 0$. For any cross section on the event horizon, a basis with the null vector fields $\{k^a, l^a, y^a\}$ can be constructed, where $l^a$ is a second null vector. Since $l^a$ and $k^a$ are both null vectors, they should satisfy the following relation
\begin{equation}
	k^a k_a = l^a l_a = 0, \qquad k^a l_a = -1\,.
\end{equation}
Using the two null vectors, the binormal of the cross section is given by $\bm{\epsilon}_{ab} = 2 k_{[a} l_{b]}$, and the induced metric on any cross section of the future event horizon is defined as
\begin{equation}
	\gamma_{ab} = g_{ab} + 2 k_{(a} l_{b)}\,.
\end{equation}
The relationship between the null vectors and the induced metric can be expressed as $k^a \gamma_{ab} = l^a \gamma_{ab} = 0$.

Since the extrinsic curvature of the event horizon $\mathcal{H}$ is defined as
\begin{equation}
	B_{ab} = \gamma_{a}^{\ c} \gamma_{b}^{\ d} \nabla_{c} k_{d}\,,
\end{equation}
the evolution of the induced metric along the future event horizon can be given as \cite{Kolekar:2012tq}
\begin{equation}\label{evolugamma}
	\gamma_{a}^{\ c} \gamma_{b}^{\ d} \mathcal{L}_{k} \gamma_{cd} = 2 \left(\sigma_{ab} + \frac{\theta}{n} \gamma_{ab} \right) = 2 B_{ab}\,,
\end{equation}
where $\sigma_{ab}$ and $\theta$ represents the shear and the expansion of the event horizon respectively. The evolution of the extrinsic curvature along the horizon can also be obtained as \cite{Gourgoulhon:2005ng}
\begin{equation}\label{excuevo}
	\gamma_{a}^{\ c} \gamma_{b}^{\ d} \mathcal{L}_{k} B_{cd} = B_{ac} B_{b}^{\ c} - \gamma_{a}^{\ c} \gamma_{b}^{\ d} R_{ecfd} k^e k^f\,.
\end{equation}
From this result, the Raychaudhuri equation can be given by
\begin{equation}
	\frac{d \theta}{d \lambda} = - \frac{\theta^2}{n} -\sigma^{ab} \sigma_{ab} - R_{kk}\,,
\end{equation}
where we have used the convention $A_{kk} = A_{ab} k^a k^b$ for any tensor $A_{ab}$.

To describe the perturbation of the dynamical fields, we introduce a sufficient small parameter $\epsilon$, which represents the order of the approximation of the perturbation. From the small parameter, we assume that $B_{ab} \sim \theta \sim \sigma_{ab} \sim \pd_\l \c\sim \pd_\l^2\c\sim \mathcal{O}(\epsilon)$, where we have denoted $\pd_\l=k^a\pd_a=\pd/\pd\l$. In the following, the symbol ``$\simeq$'' will be used to represent the identity under the linear order approximation.

Under the first-order approximation of the perturbation, the linear version of the Raychaudhuri equation can be approximately written as
\begin{equation}
	\frac{d \theta}{d \lambda} \simeq -R_{kk}\,.
\end{equation}
While Eq. (\ref{excuevo}) under the first-order approximation can also be simplified as
\begin{equation}\label{fiverbevo}
	\gamma_{a}^{\ c} \gamma_{b}^{\ d} \mathcal{L}_{k} B_{cd} \simeq - \gamma_{a}^{\ c} \gamma_{b}^{\ d} R_{ecfd} k^e k^f\,.
\end{equation}

In the following, we will examine the linearized second law of the Wald entropy for Horndeski gravity in the above dynamical geometry. The expression of the Wald entropy can be simplified as
\begin{equation}\label{Sr}
S = \frac{1}{4} \int_s d^n y \sqrt{\gamma} \left(1 + \rho\right)\,,
\end{equation}
where
\ba\begin{aligned}
\rho = - \frac{\alpha}{4} D^a \chi D_a \chi
\end{aligned}\ea
is defined as the entropy density that comes from the interaction between the gravity and the non-minimally coupled scalar field. The rate of change of the entropy along the future horizon is defined by
\begin{equation}
\frac{dS}{d \lambda} = \frac{1}{4} \int_s d^n y \sqrt{\gamma} \boldsymbol{\Theta}\,,
\end{equation}
where $\boldsymbol{\Theta}$ represents the generalized expansion of the event horizon. According to Eq. \eq{Sr}, the change rate of the entropy can also be expressed as
\begin{equation}\label{genvarenropy}
	\frac{dS}{d \lambda} = \frac{1}{4} \int_s d^n y \sqrt{\gamma} \left[\math{L}_k\rho + \theta \left( 1+ \rho \right) \right] \,,
\end{equation}
where the Lie derivative $\math{L}_k$ is commutative to the induced covariant derivative $D_a$ because the entropy density $\rho$ is an intrinsic quantity on the hypersurface $s$. For the first term on the right-hand side of Eq. (\ref{genvarenropy}), it can be calculated as
\begin{equation}\label{firstterm}
	\begin{split}
	&\int_s d^n y \sqrt{\gamma} \math{L}_k \rho  =  \frac{\alpha}{2} \int_s d^n y \sqrt{\gamma} \left[B^{ab} D_a \chi D_b \chi - D^a \chi D_a \left(\partial_\lambda \chi \right) \right] \\
		& = \frac{\alpha}{2} \int_s d^n y \sqrt{\gamma} \left[B^{ab} D_a \chi D_b \chi + \partial_\lambda \chi D^2 \chi - D_a \left(D^a \chi \partial_\lambda \chi \right)  \right]\,,
	\end{split}
\end{equation}
where we have used the definition of the extrinsic curvature in the first step. Using the Stokes' theorem, one can see that the last term only contributes a boundary term. If we assume that the cross section of the event horizon is compact, the boundary term can be neglected directly. Therefore, the result of Eq. (\ref{firstterm}) can be simplified as
\begin{equation}
	\int_s d^n y \sqrt{\gamma} \frac{d \rho}{d \lambda} = \frac{\alpha}{2} \int_s d^n y \sqrt{\gamma} \left(B^{ab} D_a \chi D_b \chi + \partial_\lambda \chi D^2 \chi \right)\,.
\end{equation}
According to the definition, one can obtain the expression of the generalized expansion $\boldsymbol{\Theta}$ as
\begin{equation}
	\boldsymbol{\Theta} = \frac{\alpha}{2} \left(B^{ab} D_a \chi D_b \chi + \partial_\lambda \chi D^2 \chi \right) + \theta \left(1 + \rho \right)\,.
\end{equation}
Under the first-order approximation, the rate of change of the generalized expansion can be expressed as
\begin{equation}\label{vartheta}
	\begin{split}
		\frac{d \boldsymbol{\Theta}}{d \lambda} & \simeq \frac{\alpha}{2} \left[\left(\mathcal{L}_k B_{ab} \right) D^a \chi D^b \chi + \partial_\lambda^2 \chi D^2 \chi \right] + \frac{d \theta}{d \lambda} \left( 1+ \rho \right) \\
		& \simeq \frac{\alpha}{2} \left[\partial_\lambda^2 \chi D^2 \chi - k^c k^d R_{acbd} D^a \chi D^b \chi \right] - (1 + \rho) R_{kk}\,,
	\end{split}
\end{equation}
where the linear version of the Raychaudhuri equation and Eq. (\ref{fiverbevo}) have been used in the last step. Following a similar considering in Ref. \cite{Bhattacharjee:2015yaa}, the rate of change of the generalized expansion can be written as
\begin{equation}\label{generalexpan}
	\frac{d \boldsymbol{\Theta}}{d \lambda} \simeq -8 \pi T_{kk} + E_{kk}\,.
\end{equation}
To investigate whether the Wald entropy of black holes in Horndeski gravity obeys the linearized second law, we just need to demonstrate that $E_{kk}$ can be neglected under the first-order approximation. It is because when considering the null energy condition $T_{kk} \ge 0$, Eq. (\ref{generalexpan}) can reduce to
\begin{equation}
	\frac{d \boldsymbol{\Theta}}{d \lambda} \le 0
\end{equation}
under the first-order approximation. Since the black hole will become a stationary state after the matter fields perturbation as mentioned above, $\boldsymbol{\Theta}$ should vanish in the asymptotic future. It implies that the value of $\boldsymbol{\Theta}$ must be positive in the future null direction, and the entropy of the black hole will increase along the future horizon under the first-order approximation. According to the above discussions, to examine whether the Wald entropy for black holes in Horndeski gravity obeys the second law, we should only demonstrate that $E_{kk}$ vanishes under the first-order approximation of the perturbation.

From Eq. (\ref{vartheta}), the specific expression of $E_{kk}$ in Horndeski gravitational theory under the first-order approximation can be expressed as
\begin{equation}\label{ekk}
	E_{kk} \simeq H_{kk} - (1 + \rho) R_{kk} + \frac{\alpha}{2} \left[\partial_\lambda^2 \chi D^2 \chi - k^c k^d R_{acbd} D^a \chi D^b \chi \right].
\end{equation}
According to the specific expression of the equation of motion, the first two terms on the right-hand side of Eq. (\ref{ekk}) can be written as
\begin{equation}\label{hrhor}
	\begin{split}
&H_{kk} - (1+\rho) R_{kk} \simeq  - \frac{\alpha}{2} \partial_\lambda^2 \chi \nabla^2 \chi + \frac{\alpha}{2} k^c k^d R_{acbd} \nabla^a \chi \nabla^b \chi \\
		& + \alpha k^b R_{bc} \nabla^c \chi \partial_\lambda \chi + \frac{\alpha}{2} k^a k^b \left(\nabla_a \nabla_c \chi \right) \left(\nabla_b \nabla^c \chi \right)\,.
	\end{split}
\end{equation}
For the first term in Eq. (\ref{hrhor}), using the definition of the induced metric on the cross section of the event horizon, $\nabla^2 \chi$ can be decomposed as
\begin{equation}\label{resultfirstterm}
	\nabla^2 \chi = -2 k^a l^b \nabla_a \nabla_b \chi + D^2 \chi\,,
\end{equation}
and the first term in the result of Eq. (\ref{hrhor}) can be expressed as
\begin{equation}
	-\frac{\alpha}{2} \partial_\lambda^2 \chi \nabla^2 \chi = - \frac{\alpha}{2} \partial_\lambda^2 \chi \left(D^2 \chi -2 k^a l^b \nabla_a \nabla_b \chi \right)\,.
\end{equation}
For the second term on the right-hand side of Eq. (\ref{hrhor}), according to the definition of the induced metric as well, the covariant derivative of the scalar fields can be decomposed as
\begin{equation}\label{covderscalarfield}
	\nabla^a \chi = -l^a \partial_\lambda \chi - k^a l^b \nabla_b \chi + D^a \chi\,.
\end{equation}
Using Eq. (\ref{covderscalarfield}), the second term under the first-order approximation can be given as
\begin{equation}\label{secondterm}
	\begin{split}
		& \frac{\alpha}{2} k^c k^d R_{acbd} \nabla^a \chi \nabla^b \chi \\
		& \simeq \frac{\alpha}{2} k^c k^d R_{acbd} D^a \chi D^b \chi - \alpha l^a k^c k^d R_{acbd} \partial_\lambda \chi D^b \chi\,.
	\end{split}
\end{equation}
In the following, we also assume that the background spacetime has a regular bifurcation surface. This additional assumption implies that the spacetime geometry on the bifurcation surface satisfy the boost symmetry, which demands that the number of $k^a$ in the expression should be equal to the number of $l^a$. For the second term on the right-hand side of Eq. (\ref{secondterm}), since $\partial_\lambda \chi$ is proportional to a first-order quantity, $l^a k^c k^d R_{acbd} \gamma_{\ e}^{b}$ should be evaluated on the background spacetime. According to the dimensional analysis, one can find that this quantity satisfy the following relation
\begin{equation}
	l^a k^c k^d R_{acbd} \gamma_{\ e}^{b} \propto \frac{1}{\lambda}\,.
\end{equation}
From the above expression, we can clearly see that this quantity will diverge at $\lambda = 0$. To ensure the regularity of the bifurcation surface in the background spacetime and satisfy the boost symmetry, the second term on the right-hand side of Eq. (\ref{secondterm}) can be neglected directly. Therefore, Eq. (\ref{secondterm}) can be simplified as
\begin{equation}\label{resultsecondterm}
	k^c k^d R_{acbd} \nabla^a \chi \nabla^b \chi \simeq k^c k^d R_{acbd} D^a \chi D^b \chi\,.
\end{equation}
Based on the decomposition of Eq. \eq{covderscalarfield}, the third term on the right-hand side of Eq. (\ref{hrhor}) can be written as
\begin{equation}\label{resultthirdterm}
	\alpha k^b R_{bc} \nabla^c \chi \partial_\lambda \chi \simeq - \alpha R_{kk} l^a \nabla_a \chi \partial_\lambda \chi + \alpha k^b R_{bc} D^c \chi \partial_\lambda \chi\simeq 0\,,
\end{equation}
under the first-order approximation, where we have used the fact $R_{kk}\sim \pd_\l\chi\sim \math{O}(\e)$, and $k^a R_{ac} \gamma^{c}_{\ b}$ vanishes in the background spacetime because it does not satisfy the requirement of the boost symmetry as well.
For the last term of (\ref{hrhor}), we have
\begin{equation}\label{lastterm}
	\begin{split}
		& \frac{\alpha}{2} k^a k^b \left(\nabla_a \nabla_c \chi \right) \left(\nabla_b \nabla^c \chi \right) = \frac{\alpha}{2} (k^a \nabla_a \nabla_c \chi ) \gamma^{cd} (k^b \nabla_b \nabla_d \chi)\\
& \quad\quad\quad\quad\quad\quad\quad\quad\quad\quad- \alpha \left(k^a k^c \nabla_a \nabla_c \chi \right) (k^b l^d \nabla_b \nabla_d \chi ) \\
		& = \frac{\alpha}{2} \left(k^a \nabla_a \nabla_c \chi \right) \gamma^{cd} (k^b \nabla_b \nabla_d \chi) - \alpha \partial_\lambda^2 \chi (k^b l^d \nabla_b \nabla_d \chi )\,.
	\end{split}
\end{equation}
For the first term in Eq. (\ref{lastterm}), we have
\ba\begin{aligned}\label{partfirstterm}
	&\left(k^a \nabla_a \nabla_c \chi\right) \gamma^{cd}\\
 &=\gamma^{cd} \nabla_c \left(\partial_\lambda \chi \right) - \gamma^{cd} \left(\nabla^a \chi  \right) \nabla_c k_a \\
		 &= \gamma^{cd} \nabla_c \left(\partial_\lambda \chi \right) - \gamma^{cd} \left(\nabla_c k_a \right) D^a \chi \\
		& + \gamma^{cd} l^a \nabla_c k_a \partial_\lambda \chi + \left(l^b \nabla_b \chi \right) k^a \left(\nabla_c k_a \right) \gamma^{cd} \\
		&= \gamma^{cd} \nabla_c \left(\partial_\lambda \chi \right) - \gamma^{cd} \left(D^a \chi\right) B_{ca} + \gamma^{cd} l^a \nabla_c k_a \partial_\lambda \chi \\
&+ \frac{1}{2}(l^b \nabla_b \chi ) D_c \left(k^a k_a \right) \gamma^{cd} \\
		&=  \gamma^{cd} \nabla_c \left(\partial_\lambda \chi \right) - \gamma^{cd} \left(D^a \chi\right) B_{ca} + \gamma^{cd} l^a \nabla_c k_a \partial_\lambda \chi \\
		&\sim \mathcal{O} \left(\epsilon\right),
\end{aligned}\ea
According to the result of Eq. (\ref{partfirstterm}), we can see that the first term in the result of Eq. (\ref{lastterm}) is a second-order quantity. Since we only consider the first-order approximation, the first term on the right-hand side of Eq. (\ref{lastterm}) can be neglected directly in our research. Therefore, the last term in the result of Eq. (\ref{hrhor}) can be reduced as
\begin{equation}\label{resultlastterm}
	\frac{\alpha}{2} k^a k^b \left(\nabla_a \nabla_c \chi \right) \left(\nabla_b \nabla^c \chi \right) \simeq - \alpha \partial_\lambda^2 \chi \left(k^b l^d \nabla_b \nabla_d \chi \right)\,.
\end{equation}
Substituting the results of Eq. (\ref{resultfirstterm}), Eq. (\ref{resultsecondterm}), Eq. (\ref{resultthirdterm}), and Eq. (\ref{resultlastterm}) into Eq. (\ref{hrhor}), we can further obtain
\begin{equation}\label{firsttwoekk}
	\begin{split}
		& H_{kk} - \left(1+ \rho\right) R_{kk} \\
		& \simeq - \frac{\alpha}{2} \partial_\lambda^2 \chi D^2 \chi + \frac{\alpha}{2} k^c k^d R_{acbd} D^a \chi D^b \chi.
	\end{split}
\end{equation}
Finally, substituting Eq. (\ref{firsttwoekk}) into Eq. (\ref{ekk}), we can obtain
\begin{equation}
	E_{kk} \simeq 0\,.
\end{equation}
According to the above discussion, this result illustrates that the Wald entropy of black holes in Horndeski gravity obeys the second law of the black hole thermodynamics at the linear order approximation.

\section{Conclusions}\label{sec4}
The entropy of stationary black holes is modified by adding a scalar field term in Horndeski gravity because of the existence of non-minimally coupled scalar field, where the expression of the entropy is given by the Wald entropy. Based on it, an issue of whether the Wald entropy of black holes in Horndeski gravitational theory can still obey the linearized second law of black hole thermodynamics is naturally proposed. In order to investigate this issue, a physical process that the black hole in Horndeski gravity is perturbed by the accreting matter fields and finally settles down to a stationary state is considered. Subsequently, two assumptions are suggested, which state that the matter fields always satisfy the null energy condition, and there is a regular bifurcation surface in the background spacetime. According to the Raychaudhuri equation, we demonstrate that the change rate of the generalized expansion along the future horizon is negative. Since the final state of the black hole after the perturbation is demanded as a stationary state, the value of the generalized expansion must be positive according to the result. It indicates that the entropy of the black holes in Horndeski gravitational theory continuously increases along the future event horizon, and the Wald entropy of black holes in Horndeski gravity satisfies the linearized second law. This result reinforces the physical explanation of the Wald entropy in Horndeski gravity and takes a step towards discussing the second law of thermodynamics in the gravitational theories with non-minimal coupling matter fields.

\section*{Acknowledgement}
This research was supported by National Natural Science Foundation of China (NSFC) with Grants No. 11775022 and 11873044.

\end{document}